\documentclass[referee]{raa}            

\usepackage{graphicx,times}             
\input{epsf.sty}                        
\input{psfig.sty}                       

\begin{document}

   \title{Study of Temporal Evolution of Emission Spectrum in a Steeply Rising Submillimeter Burst
}

   \volnopage{Vol.0 (200x) No.0, 000--000}      
   \setcounter{page}{1}          

   \author{J. P. Li
      \inst{1,2}
   \and A. H. Zhou
      \inst{1,2}
   \and X. D. Wang
      \inst{3}
   }

   \institute{Key Laboratory of Dark Matter and Space Astronomy, Chinese Academy of Sciences; {\it zhouah@pmo.ac.cn}\\
        \and
              Purple Mountain Observatory, Chinese Academy of Sciences, Nanjing 210008\\
        \and
             Hohai University, Nanjing 210098, China\\
   }

   \date{Received~~2015 month day; accepted~~2015~~month day}

\abstract{In the paper the spectral temporal evolution of a steeply rising submillimeter (THz) burst occurred on 2003 November 2 was investigated in detail for the first time. Observations show that the flux density of the THz spectrum increased steeply with frequency above 200 GHz. Their average rising rates reached a value of 235 sfu/GHz (corresponding spectral index $\alpha$ of 4.8) during the burst. The flux densities reached about 4,000 and 70,000 sfu at 212 and 405 GHz at maximum phase, respectively. The emissions at 405 GHz maintained continuous high level that they exceed largely the peak values of the microwave (MW) spectra during the main phase. Our studies suggest that only energetic electrons with a low-energy cutoff of $\sim$1 MeV and number density of $\sim$$10^{6}$--$10^{8}$ cm$^{-3}$ can produce such strong and steeply rising THz component via gyrosynchrotron (GS) radiation based on numerical simulations of burst spectra in the nonuniform magnetic field case. The electron number density $\emph{N}$, derived from  our numerical fits to the THz temporal evolution spectra, increased substantially from $8\times10^{6}$  to $4\times10^{8}$ cm$^{-3}$, i.e., $\emph{N}$ value increased 50 times during the rise phase. During the decay phase it decreased to $7\times10^{7}$ cm$^{-3}$, i.e., decreased about five times from the maximum phase. The total electron number decreased an order of magnitude from the maximum phase to the decay phase. Nevertheless the variation amplitude of $\emph{N}$ is only about one time in the MW emission source during this burst, and the total electron number did not decrease but increased by about 20$\%$ during the decay phase. Interestingly, we find that the THz source radius decreased by about 24$\%$ while the MW source one, on the contrary, increased by 28$\%$ during the decay phase.
}
   \authorrunning{J. P. Li, A. H. Zhou \& X. D. Wang }            
   \titlerunning{Study of Temporal Evolution of Emission Spectrum}  

   \maketitle

%
%
\section{Introduction}           
\label{sect:intro}
The diagnostics of high relativistic electrons and emission source regions in solar flares are essential based on their emission spectra at various wavelength ranges (Gary 1985; Wang et al. 1994; Zhou and Karlicky 1994; Zhou et al. 2005; Huang et al. 2005; Zhou et al. 2009). The spectral maximum of the MW emission is typically in the range of 3--30 GHz, depending primarily on the energy of the accelerated electrons. There are rare microwave spectral examples peaking at higher frequencies, up to 94 GHz for solar observations made in the past (Croom 1973; Kaufmann et al. 1985; Ramaty et al. 1994; Chetok et al. 1995). Since 2000 new instrumentation observing in the 200--400 GHz range became available, more than 10 flares have been observed in this band (Silva et al. 2007; Krucker et al. 2013). For some flares these observations show that the gyrosynchrotron (GS) component extends up to 200 GHz (Trotter et al. 2002) or higher frequencies (l{\"u}thi et al. 2004b). However, for other flares the radio spectrum above 200 GHz is not the continuation of the GS spectrum measured at lower frequencies, but surprisingly increases with increasing frequency (Silva et al. 2007; l{\"u}thi et al. 2004a; Kaufmann et al. 2004). This spectral feature is termed a ``THz component''.

So far, there are various possible explanations proposed for the new increasing submillimeter spectral component. But many theoretical issues remain open (Zhou et al. 2011; Krucker et al. 2013). We think that the GS emission is a possible mechanism of the THz emission. But under the uniform source assumption, an extreme value of magnetic field of 4500 G and a high electron number density of $1.7\times10^{12}$ cm$^{-3}$ are required in GS emission explanation (Silva et al. 2007). These requirements can be decreased to a reasonable range by using our GS radiation model including the self and gyroresonance absorptions (Zhou et al. 2008) in a nonuniform magnetic field model (Zhou et al. 2011). 

Thus far, the positive-slope spectra have been observed only in a handful of the most energetic events (Krucker et al. 2013). So the THz burst observations are very valuable, especially for the 2003 November 2 burst which has a set of complete THz spectral temporal evolution observations. These observations can provide important diagnostics about energy release process of ultra-relativistic electrons and environment variation of THz burst region in deeper solar atmosphere layers (about 1000--30000 km above the photosphere).  

In the paper, we will investigate this set of temporal evolution spectra in detail for the first time. A vast amount of numerical simulations for these temporal evolution spectra have been done by using GS emission model in the magnetic dipole field case. We try to obtain diagnostic of physical parameters of high relativistic electrons  and their environment in the solar THz and MW burst regions. Then we compare the temporal evolution results about electron number density and source size in the MW and THz burst regions. Finally we give summary and conclusions.


 \section{Observations}
\label{sect:Obs}
Extensive flare activities were observed in super-AR NOAA 10486 during its disk passage (October 22 -- November 4, 2003). Among them an increasing submillimeter burst was detected by Solar Submillimeter Telescope (SST) at 212 and 405 GHz in the flare starting at $\sim$17:16 UT on November 2, 2003. The flare was also detected simultaneously by the Owens Valley Solar Array (OVSA) at MW wavelengths (see Figure 1). The maximum phase, main phase, and decay phase are shown in Figure 1 according to the time profiles of the THz burst. This flare was classified as a GOES X8.3 and 2B event. Figure 2 gives the temporal evolutions of emission spectrum at the MW and THz wavelengths. It shows that at 17:16:15 UT of the rise phase, the  OVSA radio flux densities reached, respectively, $5.1\times10^{3}$ and $3\times10^{4}$ sfu at 3 and 18 GHz of the MW range; and $\sim$$1.2\times10^{3}$ and $3.1\times10^{4}$ sfu at 212 and 405 GHz of the THz range. At the maximum phase the flux densities increased dramatically to $8\times10^{3}$ and $4\times10^{4}$ at 3 and 18 GHz; and $4\times10^{3}$ and $7\times10^{4}$ sfu at 212 and 405 GHz. After the maximum phase the flux densities gradually decreased until 17:18:00 UT. In a period of 17:18:00 to 17:18 30 UT the flux densities of the submillimeter spectrum increased once again, while this increase of flux density did not occur in the MW range, which means that more energetic electrons were accelerated to higher energies in that period. After $\sim$17:20:30 UT the MW and THz emissions decreased further. The flux densities at 405 GHz are so high that they exceed the peak flux densities of the MW spectrum during the main phase. 

\begin{figure}
	\centering
	\includegraphics[width=0.8\textwidth, angle=0]{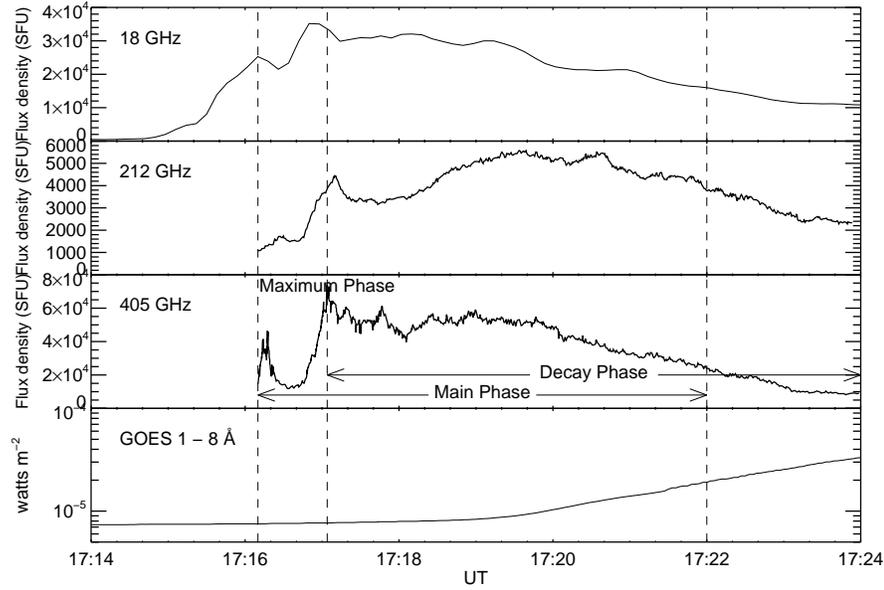}
	\caption{Temporal evolution of the emission at 18 GHz from OVSA, 212 and 405 GHz from SST, and GOES X-ray flux of the 2003 November 2 flare. The rise phase ($\sim$17:16 -- $\sim$ 17:17 UT), maximum one ($\sim$17:17 UT), decay one ($\sim$17:17 -- $\sim$17:24 UT, and  main one ($\sim$17:16 --17:22 UT) are shown in panel three.}\label{FIG:fig1}
\end{figure}

\begin{figure}
   	\centering
   	\includegraphics[width=0.8\textwidth, angle=0]{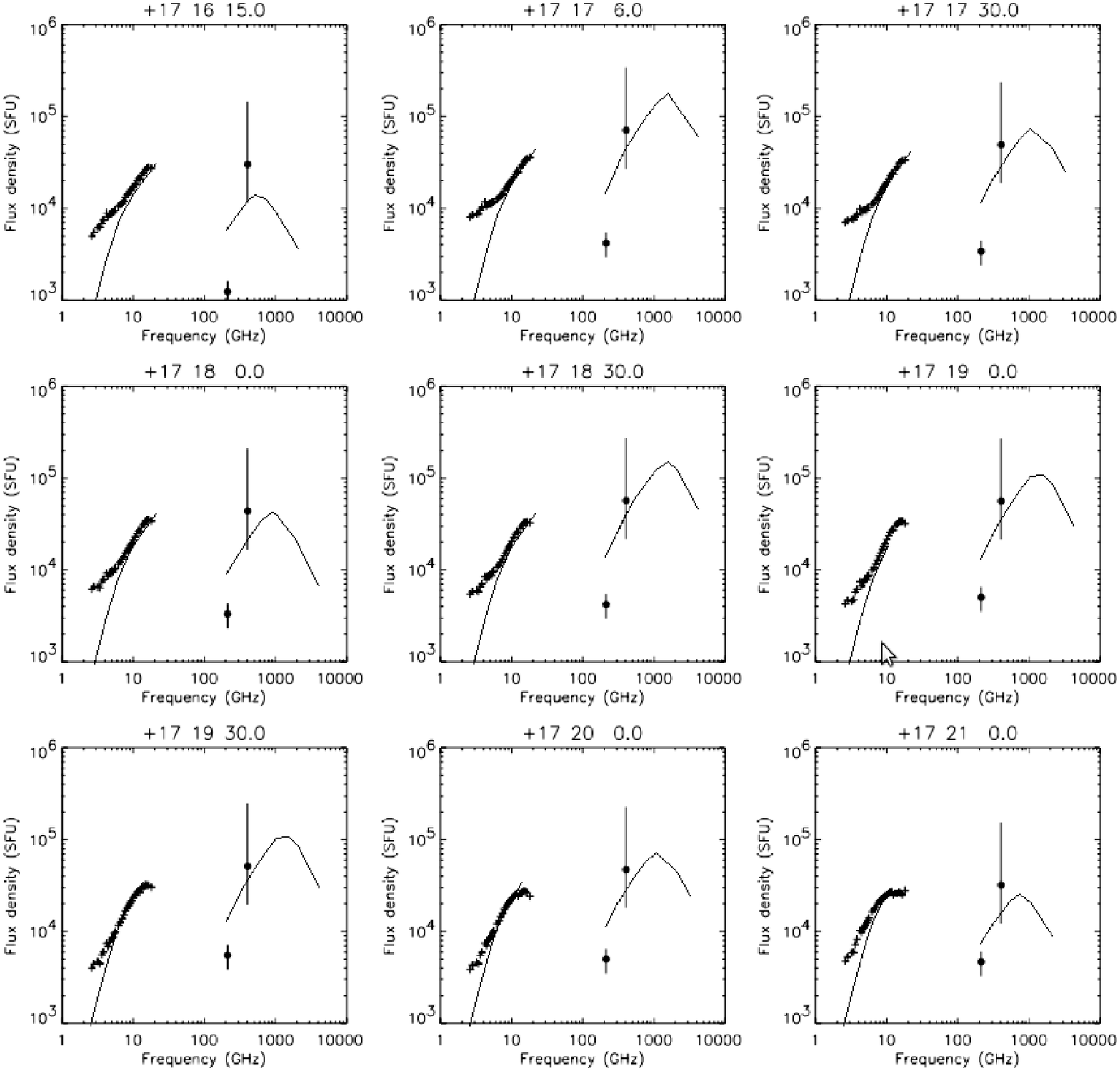}
   	\caption{The temporal evolutions of
   		radio spectrum from 17:16:15 to 17:21:00 UT of the 2003 November 2 burst
   		given by Silva et al. (2007) and for their fits (see the solid
   		lines).}\label{FIG:fig2}
\end{figure}

\section{INCREASING RATE OF FLUX DENSITY OF SUBMILLIMETER BURST SPECTRUM}
\label{sect:incr}
It is found from Figure 2 that the rising rate of the flux density, $\emph{r}$ (sfu/GHz), changes greatly in the THz spectrum range during the burst. The rising rate obtained from the observational spectra increased from 154 to 342 sfu/GHz during the rise phase (see Table 1). During the decay phase it decreased from 342 to 142 sfu/GHz, i.e., the rising rate at the maximum phase is much higher than that at the rise phase and decay phase. The average value of $\emph{r}$ reached 235 sfu/GHz (corresponding spectral index $\alpha$ of 4.8 at the optically thick part) during this flare. Thus, it is a steeply rising THz burst.

\begin{table*}
	\begin{center}
		\caption{Rising Rates $\emph{r}$ sfu/GHz of the Flux Density $S_{\nu}$ of the THz Component During the November 2 THz Burst.}\label{TAB:tab1}
		\begin{tabular}{ccp{0.8cm}p{0.8cm}p{0.8cm}cp{1.2cm}c}
			\hline\noalign{\smallskip}
			date      &  time &  Rise-phase   & Max.-phase    & Decay-phase &$S_{212GHz} $&$S_{405GHz}$ &$r(sfu/GHz)$\\
			\hline\noalign{\smallskip}
			2003 11 02&$17:16:15$&$  yes       $&$            $&$            $&$1.2\times10^{3}$&$ 3.1\times10^{4}$&$154$\\
			&$17:17:06$&$            $&$ yes        $&$            $&$4.0\times10^{3}$&$ 7.0\times10^{4}$&$342$\\
			&$17:17:30$&$            $&$            $&$ yes        $&$3.2\times10^{3}$&$ 5.0\times10^{4}$&$242$\\
			&$17:18:00$&$            $&$            $&$ yes        $&$3.5\times10^{3}$&$ 4.0\times10^{4}$&$210$\\
			&$17:18:30$&$            $&$            $&$ yes        $&$4.0\times10^{3}$&$ 5.8\times10^{4}$&$280$\\
			&$17:19:00$&$            $&$            $&$ yes        $&$5.0\times10^{3}$&$ 5.5\times10^{4}$&$259$\\
			&$17:19:30$&$            $&$            $&$ yes        $&$5.0\times10^{3}$&$ 5.5\times10^{4}$&$259$\\
			&$17:20:00$&$            $&$            $&$ yes        $&$5.0\times10^{3}$&$ 4.8\times10^{4}$&$223$\\
			&$17:21:00$&$            $&$            $&$ yes        $&$4.5\times10^{3}$&$ 3.2\times10^{4}$&$142$\\
			
			\hline
		\end{tabular}
	\end{center}
\end{table*}

\section{FIT FOR THE STEEPLY RISING SUBMILLIMETER  BURST SPECTRUM}
\label{sect:fits}

It is well known that radio spectrum can provide crucial information about energetic electrons and their environment in solar flares. These information contains mainly the energy spectral index $\delta$, low- and high-energy cutoffs $E_{0}$ and $E_{m}$, electron number density $\emph{N}$, source radius $R$, and magnetic field strength $B$ in source regions. The magnetic field strength $B$ can been estimated in the case of magnetic dipole field if the photosphere magnetic field strength $B_{0}$, the lower boundary height of THz source $h_{d}$, and the upper one $h_{u}$ can be determined (Zhou et al. 2008, 2011). A set of reasonable values of the parameters is taken as $\delta$ = 3,  $B_{0}$ = 5000 G, $R = 0.\arcsec5$, $h_{d}$ = $10^{8}$~cm, and $h_{u}$ = $3\times10^{9}$~cm for the THz spectra based on our numerical simulations. So only the energy cutoffs $E_{0}$, $E_{m}$, and electron number density $\emph{N}$ remain unknown. Figure 3 shows that the effect of increasing high-energy cutoff $E_{m}$ on the THz spectrum, it only enhances the GS emission at the optically thin part a little and at the optically thick part it is constant. So the $E_{m}$ value of 10 MeV is high enough for the THz  burst spectrum calculations. Finally the remained unknown parameters are only $E_{0}$ and $\emph{N}$. In the paper we try to get the two parameters from the fits to THz spectral evolution observations.
  
  \begin{figure}
  	\centering
  	\includegraphics[width=0.8\textwidth, angle=0]{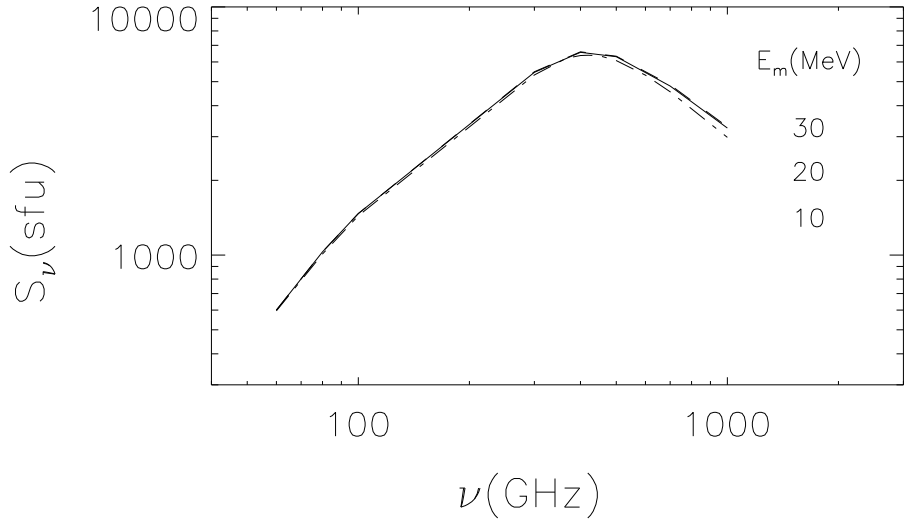}
  	\caption{Calculated GS emission
  		spectra in the THz range in the magnetic dipole field case for a
  		sequence of the high-energy cutoffs $E_{m}$, where $\delta = 3$,
  		$E_{0} = 500$~keV, $N = 10^{7}$~cm$^{-3}$, $\theta = 60^\circ$, and
  		$B_{0} = 5000$~G.}
  	\label{Fig:fig3}
  \end{figure}

Figure 2 shows that their flux densities reached about 4,000 and 70,000 sfu at 212 and 405 GHz at the maximum phase, respectively. During the main phase, the emissions at 405 GHz maintained continuous high level that they exceed largely the peak values of the microwave spectra. It is important to know what conditions are needed to produce such steeply rising and giant submillimeter emission.

Table 2 gives theoretical rising rate  $\emph{r}_{theo}$ sfu/GHz of the modeled submillimeter components for different number densities $\emph{N}$ setting $E_{0}=1$ MeV, and different low-energy cutoffs $E_{0}$ setting $N = 8\times10^{11}$ cm$^{-3}$. It shows that the theoretical rising rate $r_{theo}$ increases obviously with increasing number density and increasing low-energy cutoff. The effects of low-energy cutoff and electron number density on the THz spectrum is given in Figure 4. It shows that the THz spectral distributions are sensitive to the two parameters.  We find that only the electrons with a low-energy cutoff of $\sim$1~MeV and number density of  $\sim$$10^{6}$--$10^{8}$ cm$^{-3}$ can produce such steeply rising THz spectral components as the 2003 November 2 burst. Even in this case of 1 MeV low-energy cutoff and $8\times10^{11}$ cm$^{-3}$ electron number density, the maximum theoretical rising rate $r_{theo}$ reaches only 234 sfu/GHz (see Table 2), which is  still smaller than the observational one (342 sfu/GHz) at the maximum phase of this THz burst.

\begin{figure}
	\plottwo{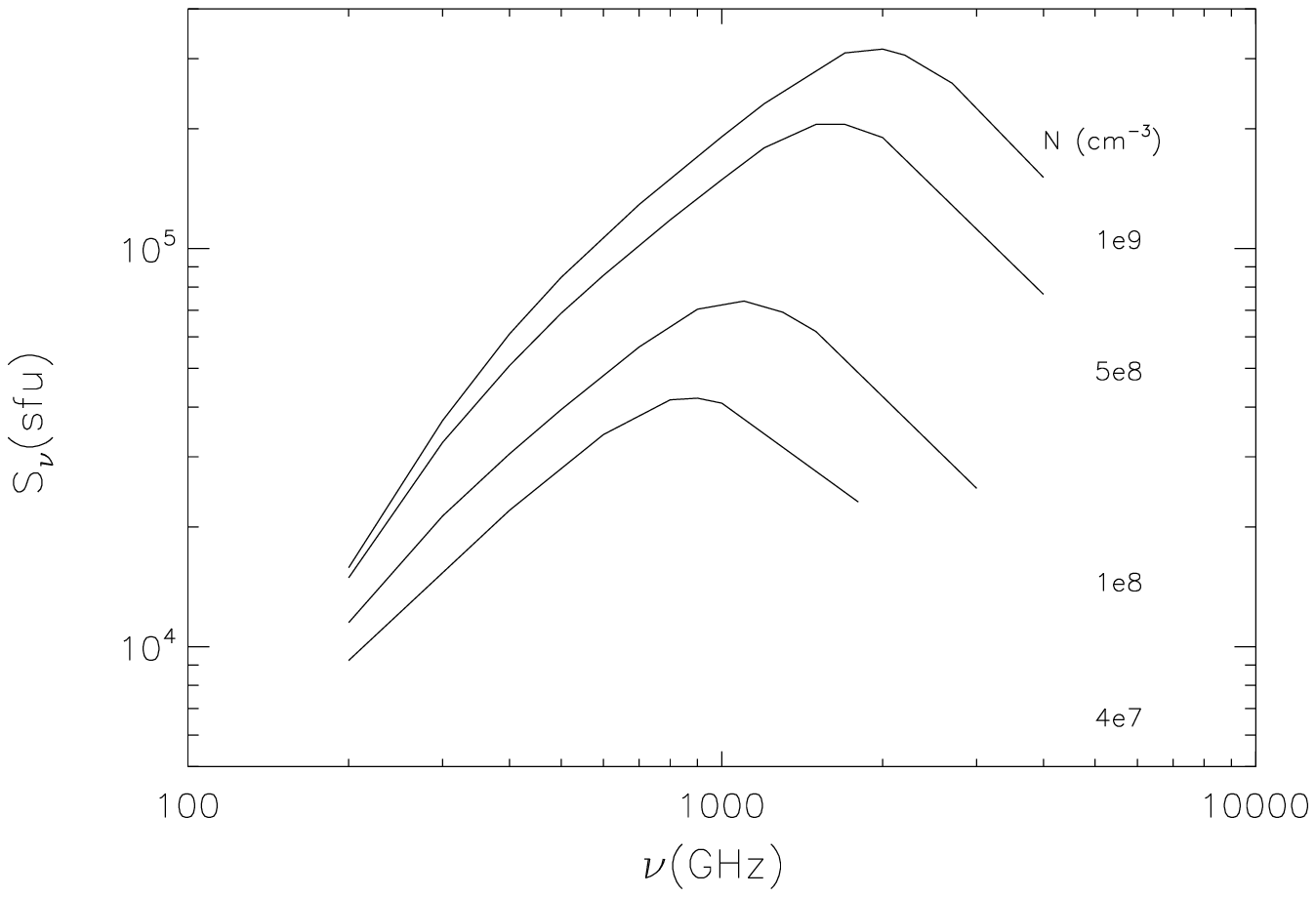} {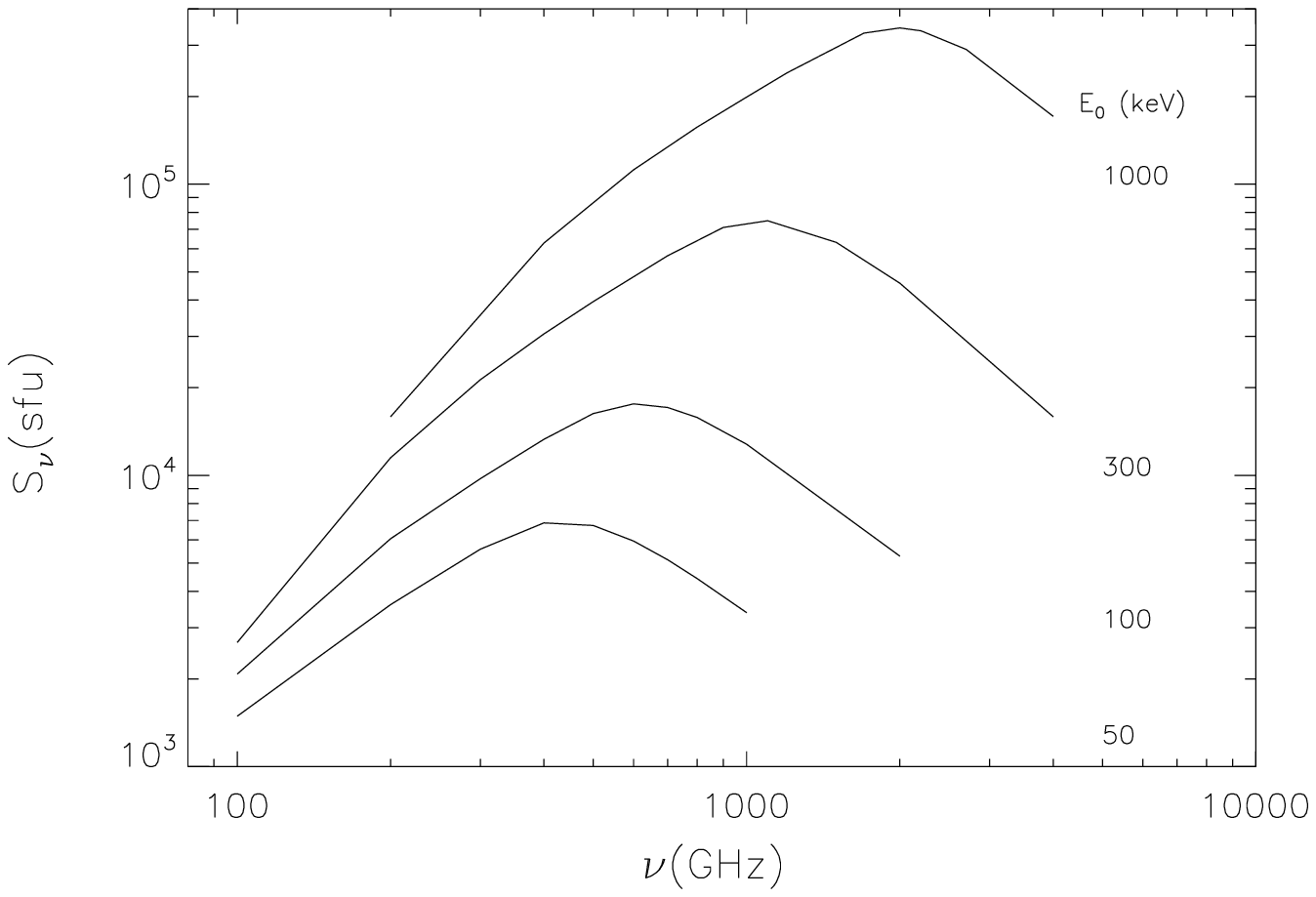}
	\caption{Calculated GS 	emission spectra in the THz range in the magnetic dipole field case for a sequence of low-energy cutoffs $E_{0}$ setting $N = 8\times10^{11}$~cm$^{-3}$  (left panel) and for  a sequence of electron number densities $\emph{N}$ setting $E_{0}=500$~keV (right panel), where $\delta = 3$, $B_{0} = 5000$~G, $\theta = 60^\circ$, and
		$h_{d} = 10^{8}$~cm.}
	\label{Fig:fig4}
\end{figure}

It is also seen from Figure 4 that the maximum frequency in the THz range can reach as high as $\sim$2000~ GHz and the GS emissions can extend to  higher ($>$ 5000 GHz)  frequencies in the case of 1 MeV low-energy cutoff and $8\times10^{11}$ cm$^{-3}$ electron number density.

\begin{table}[ht]
	\begin{center}
		\caption{Theoretical Increasing Rates $r_{theo}$ sfu/GHz of the
			Submillimeter Spectral Components for Different Number Densities N
			Setting $E_{0} = 1$ MeV, and for Different Low-Energy Cutoffs $E_{0}$ Setting $N = 8\times10^{11}$~cm$^{-3}$. Where $\delta = 3$,
			$B_{0} = 5000$~G,  $\theta = 60^\circ$, and
			$h_{d} = 10^{8}$~cm.}\label{TAB:tab2}
		\begin{tabular}{ccccc}
			\hline\noalign{\smallskip} \hline\noalign{\smallskip}
			$N$ (cm$^{-3})$&$4\times10^{7}$ &$10^{8}$&$5\times10^{8}$ &$10^{9}$\\
			$r_{theo} $&$ 64$ &$95$&$180$ &$226$\\
			$E_{0}$ (keV)&$50$ &$100$&$300$ &$1000$\\
			$r_{theo} $&$ 16$ &$36$&$96$ &$234$\\
			
			\hline
		\end{tabular}
	\end{center}
\end{table}

Now we will try to fit the temporal evolution spectra of the THz burst for $\delta = 3$, $E_{0} = 1$~MeV, and source radius $R = 0.\arcsec5$. A sequence of number densities are selected to fit these spectra. The modeled GS emission spectra are given in Figure 2 by the solid lines. It shows that the modeled spectra fit well the observational ones. The required electron  number densities are given in Table 3 for the THz spectra. It shows that the number density $\emph{N}$ increased substantially from $8\times10^{6}$~cm$^{-3}$ at the rise phase to $4\times10^{8}$~cm$^{-3}$ at the maximum phase in the THz source,  i.e., $\emph{N}$ increased about 50 times. Then the electron number number $\emph{N}$  began to drop from the maximum value, but it increased once again at 17:18:30 UT and the peak frequency of the modeled spectrum can shift to higher frequency of 1500 GHz at that time (see Figure 2). At 17:21:00 UT of the decay phase the value of $\emph{N}$  decreased to $7\times10^{7}$~cm$^{-3}$, i.e., it decreased nearly five  times from the maximum value. The total electron numbers $N_{total}$ are also calculated in radio sources (see Table 3). We can see from Table 3 that the value of  $N_{total}$ in the THz source increased rapidly from $10^{31}$ at the rise phase to $5.2\times10^{32}$ at the maximum phase, i.e., also increased 50 times as the electron number density due to constant source size in that period. During the decay phase $N_{total}$ value decreased from $5.2\times10^{32}$ to $5.3\times10^{31}$, i.e., it decreased about an order of magnitude.

We also set a sequence of electron number densities to fit  the observational MW spectra in the case of $E_{0} = 10$~keV, and $R=25\arcsec$. We found that these modeled spectra also fit well the observational spectra  from the rise phase to  decay phase. The required electron number density $\emph{N}$ in the MW source are also given in Table 3. It shows that the value of $\emph{N}$ in the MW emission source increased only about one time during  the rise phase and $\emph{N}$ decreased only a little during decay phase, which are much smaller than that in the THz  emission source. In the MW source the total electron number increased one time during the rise phase. During the decay phase $N_{total}$ did not decrease but increased from $5.9\times10^{35}$ at the maximum phase to $7\times10^{35}$, i.e., it increased by about 20$\%$. 

\begin{table*}
	\begin{center}
	\caption{Variations of the Source Size $R\arcsec$ and the Electron Number Density $\emph{N}$, and the Total Number Ddensity $\emph{N}_{total}$ in the MW and THz Emission Regions of the 2003 November 2 Burst.}\label{TAB:tab3}
	\begin{tabular}{ccccccc}
		\hline\noalign{\smallskip}
		Time & MW:~$R\arcsec$& $N$~(cm$^{-3}$)&${N_{total}}$& THz:~$R\arcsec$&$N$~(cm$^{-3}$)& ${N_{total}}$\\
		\hline\noalign{\smallskip}
		$17:16:15$&$25 $  &$8.0\times10^{7}$&$2.6\times10^{35}$&$0.5    $&$8\times10^{6}$&$1.0\times10^{31}$\\
		$17:17:06$&$25   $&$1.8\times10^{8}$&$5.9\times10^{35}$&$0.5    $&$4\times10^{8}$&$5.2\times10^{32}$\\
		$17:17:30$&$25   $&$1.6\times10^{8}$&$5.3\times10^{35}$&$0.5    $&$ 1.0\times10^{8}    $&$1.3\times10^{32}$\\
		$17:18:00$&$25   $&$1.6\times10^{8}$&$5.3\times10^{35}$&$0.5    $&$4\times10^{7}$&$5.2\times10^{31}$\\
		$17:18:30$&$25   $&$1.6\times10^{8}$&$5.3\times10^{35}$&$0.5    $&$3\times10^{8}$&$3.9\times10^{32}$\\
		$17:19:00$&$25   $&$1.5\times10^{8}$&$5.0\times10^{35}$&$0.5    $&$2\times10^{8}$&$2.6\times10^{32}$\\
		$17:19:30$&$30   $&$1.3\times10^{8}$&$6.1\times10^{35}$&$0.45   $&$2\times10^{8}$&$2.2\times10^{32}$\\
		$17:20:00$&$30   $&$1.3\times10^{8}$&$6.1\times10^{35}$&$0.45   $&$1.3\times10^{8}$&$1.4\times10^{32}$\\
		$17:21:00$&$32   $&$1.3\times10^{8}$&$7.0\times10^{35}$&$0.38   $&$7\times10^{7}$&$5.3\times10^{31}$\\
		\hline
	\end{tabular}
	\end{center}
\end{table*}

Here we must point out that in the period of 17:19:30 to 17:21:00 UT of the decay phase,  smaller radii ($0.\arcsec45$  and $0.\arcsec$38) for the THz source are selected to fit these  observational spectra. The effect of source size on the THz spectrum can be seen from Figure 5. It shows that the rising rate of flux density at the optically thick part of the THz spectrum increases with decreasing source size and the peak of the modeled spectrum shifts to higher frequency under a constant electron total number conditions.  If we still take the same source size ($R=0.\arcsec5$), then the required electron number $\emph{N}$ will decrease largely, which leads to the modeled flux densities of the GS emission at 405 GHz being always lower than the observations. Interestedly, we also find that the MW source  radius obtained from the  spectrum fit increased from 25$\arcsec$ to 32$\arcsec$ during the decay phase, i.e., the MW source radius increased by 28$\%$.

\begin{figure}
	\includegraphics[width=0.8\textwidth, angle=0]{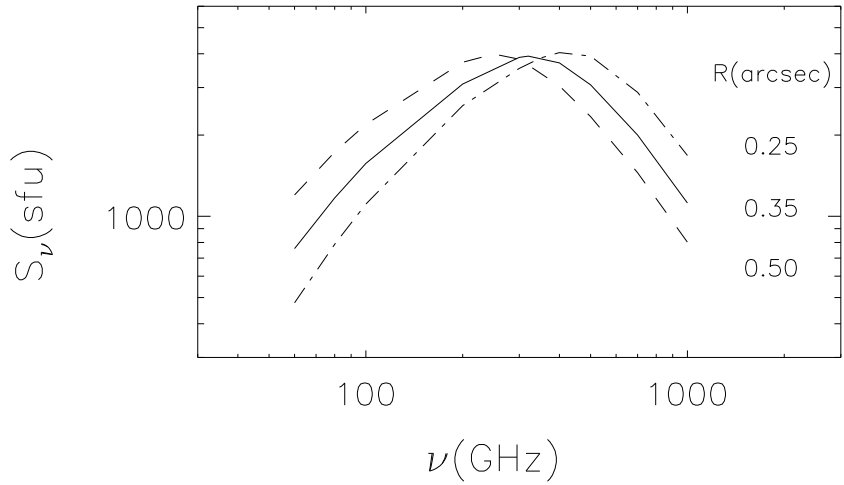}
	\caption{The effect of the emission source radius $R$ on the modeled GS emission spectrum in the THz range under a constant electron total number condition, where $\delta = 2.2$, $E_{0} = 1.5$~MeV, $B_{0} = 5000$~G, and
		$\theta = 10^\circ$. }
	\label{Fig:fig5}
\end{figure}

\section{Summary and Conclusions}  
\label{sect:fits}
Table 3 shows that the required electron number density $\emph{N}$ in the 2003 November 2 THz emission source increased substantially from $8\times10^{6}$ at the rise phase to $4\times10^{8}$~cm$^{-3}$ at the maximum phase, i.e., $\emph{N}$ value increased 50 times from the rise phase to the maximum phase. It means that there would be a  very effective electron acceleration mechanism, which can accelerate effectively a huge amount of electrons to higher energy range of $\sim$ 1 to 10~ MeV in that period. Then $\emph{N}$ began to drop from the maximum value, but it increased once again at  17:18:30 UT and the peak frequency of the fitting spectrum at that  time can shift to higher frequency of 1500 GHz (see Figure 2). It  could result from another effective electron acceleration before $\sim$17:18:30 UT. At 17:21:00 UT it decreased to $7\times10^{7}$~cm$^{-3}$, i.e., decreased about five times from the  maximum phase to the decay phase. But the variation amplitude of $\emph{N}$ in the MW source reached only about one time during this burst, which is much smaller than that in the THz emission source. During the decay phase the total electron number $N_{total}$ decreased an order of magnitude in the THz source, while  the value of $N_{total}$ in the MW source did not decrease but  increased  from $5.9\times10^{35}$ to $7\times10^{35}$ , i.e., it increased by about 20$\%$. 

The dramatic variation of electron number density in the THz emission source could result from the effective electron acceleration at the rise phase and strong electron energy loss at the decay phase. While in the MW source,  $N_{total}$ did not decrease but  increased by about 20$\%$ during the decay phase. There could be much more electrons decayed from the higher energies and lower electron energy loss, so the variation amplitude of electron number density in the MW source is much smaller than that in the THz source.

It is found that the THz source radius obtained from numerical fits decreased from $0.\arcsec5$ to $0.\arcsec45$ even to $0.\arcsec38$  during the decay phase, i.e., it decreased by about 24$\%$, but the MW one increased by 28$\%$  during the decay phase. Similar variation of source size can also be seen from the study of the 2003 November 4 event (Zhou et al. 2011). This source size variation maybe is a rather interesting result. It would  result from the trap height variation of the energetic electrons, the magnetic field topology variation, or others.

In the paper we investigate the novel rising THz burst occurred in Active Region NOAA 10486 on 2003 November 2. Our studies show that it is a steeply rising and very giant THz event. The average rising rate of the flux density reaches 235 sfu/GHz ($\alpha=4.8$) for the THz burst. The steeply rising THz spectrum can be produced by energetic electrons with a low-energy cutoff of $\sim$1~ MeV and number density of $\sim$$10^{6}$--$10^{8}$ cm$^{-3}$ in a compact source ($\sim$ 0.$\arcsec5$ in radius ) with strong local magnetic fields varying from 4590 to 780 Gauss (2690 G in mean value) via the GS emission in magnetic dipole field case. The average magnetic filed strength of 2690 G is much lower than that one (4500 G) obtained in the uniform magnetic field assumption (see Silva et al. 2007). The photosphere magnetic field of 5000 G would be possible on observation in a compact source. Unfortunately, because this active region, AR 10486, was close to the limb on November 2, there is no accurate measurement of the photosphere magnetic field. However a magnetic field of 4200 G was measured for AR 10484 (Mt Wilson sunspot group 31909) on 2003 October 22. Since region 10486 was more active than 10484, in fact producing the largest flare on record two days later on November 4, it probably had a very high magnetic field at the time of the November 2 flare, even higher than 4200 G. In fact, Livingston et al. (2006) report that 0.2$\%$ of almost 32000 active region studied have sunspots with magnetic fields larger than 4000 G, even as high as 6100 G, the highest magnetic field ever measured (see Silva et al. 2007). The associated microwave spectral components can be produced by energetic electrons with a 10 keV low-energy cutoff and a mean local magnetic field strength in an extend source of 25$\arcsec$ -- 32$\arcsec$ radius.

It is found from the numerical simulations for  the temporal evolution spectra that the variation amplitudes of electron number density and total electron number are much larger in the THz emission source than that in the MW source during the burst. In addition, the THz source radius decreased by about 24$\%$, but the MW one increased by 28$\%$ during the decay phase. These interesting results would be significant, because they can provide important information about the ultrarelativistic electron acceleration, trap, energy loss, and a possible magnetic field topology evolutions at different levels in the burst source region or others. However we must note that the required source radius is usually much small based on the GS emission calculations. Further progress in understanding the physics of THz emission requires observations with a more complete spectral coverage and higher spatial resolution at the THz range.

\begin{acknowledgements}
	The authors thank  Dr. V. F. Melnikov for fruitful discussions. This
	study is supported by the NFSC project with No.11333009 and the ``973'' program with No. 2014CB744200.
\end{acknowledgements}

\label{lastpage}

\end{document}